\documentclass[a4paper,english]{elsarticle}
\usepackage{booktabs}
\usepackage{amsmath}
\usepackage{amssymb}
\usepackage{graphicx}
\usepackage{upgreek}
\usepackage{url}
\usepackage{lineno}
\usepackage{a4wide}
\usepackage{subcaption}

\modulolinenumbers[1]

\begin{document}

\begin{frontmatter}

\title{Analysis methods for highly radiation-damaged SiPMs
\tnoteref{mytitlenote}}
\tnotetext[mytitlenote]{Talk presented at the 15$^{th}$ Vienna Conference on Instrumentation,  Vienna, Feb. 18$^{th}$--22$^{nd}$, 2019.}

\author{S. Cerioli}
\author{E. Garutti}
\author{R. Klanner\corref{Corr}}
\author{S. Martens\corref{NoCorr}}
\author{J. Schwandt}
\author{M. Zvolsk\'y\corref{Corr1}}
\address{Institute for Experimental Physics, Hamburg University}
\cortext[Corr]{Corresponding author, robert.klanner@desy.de.}
\cortext[Corr1]{now at the Institut f\"ur Medizintechnik, Universit\"at zu L\"ubeck.}

\renewcommand{\thefootnote}{\fnsymbol{footnote}}






\begin{abstract}
 Prototype SiPMs with 4384 pixels of dimensions $15 \times 15~\mu $m$^2$ produced by KETEK have been irradiated with reactor neutrons to eight fluences between $10^9$ and $5\times 10^{14}$~cm$^{-2}$.
 For temperatures between $-30~^\circ $C and $+30~^\circ $C capacitance--voltage, admittance--frequency, current--forward voltage, current--reverse voltage and charge--voltage measurements with and without illumination by a sub-nanosecond laser have been performed.
 The data have been analysed using different methods in order to extract the dependence on neutron fluence and temperature of the electrical parameters, the breakdown voltage, the activation energy for the current generation, the dark-count rate and the response to light pulses.
 The results from the different analysis methods are compared.
\end{abstract}


\end{frontmatter}


  \tableofcontents

 \section{Introduction}
  \label{Introduction}


 Since about 10 years silicon photo-multipliers (SiPMs) are becoming the photo-detectors of choice for many applications in research, medicine and industry.
 One major limitation of SiPMs
 is radiation damage.
  In this contribution methods for characterising neutron-damaged SiPMs are developed and demonstrated for a specific SiPM.
 The questions addressed are:
 \begin{itemize}

   \item Which SiPM parameters are affected by radiation damage?
   \item Which are the best methods to characterise radiation-damaged SiPMs?
   \item What are the optimal operating conditions for radiation-damaged SiPMs?

 \end{itemize}

 For the study prototype SiPMs with 4384 pixels of dimensions $15 \times 15~\mu $m$^2$ produced by KETEK have been irradiated by reactor neutrons to eight fluences between $10^9$ and $5\times 10^{14}$~cm$^{-2}$.
 These SiPMs have been previously characterised in detail before irradiation\,\cite{Chmill:2017,Chmill1:2017}, and first results after irradiation are given in Ref.~\cite{Vignali:2017}.

 For temperatures between $-30~^\circ $C and $+30~^\circ $C capacitance--voltage, admittance--frequency, current--forward voltage, current--reverse voltage and charge--voltage measurements with and without illumination by a sub-nanosecond laser have been performed.
 Different methods are used to extract  different SiPM parameters and their change with neutron fluence.
 If a parameter can be determined in different ways, the results are compared and recommendations given, which method appears to be more reliable.
 Note that only a single type of SiPM has been investigated, and some of the methods developed may not be applicable for SiPMs of different design.

 \section{Measurements and results}
  \label{sect:Results}

  The most dramatic effect of radiation damage is the increase of the dark-count rate, $DCR$, and of the dark current of the SiPM, $I_{dark}$, by many orders of magnitude~\cite{Garutti:2018}.
  Fig.~1 shows current transients of the SiPM studied before and after irradiation to a fluence $\Phi = 10^{13}$~cm$^{-2}$.
  Whereas for $\Phi = 0$ signals from single Geiger discharges can be easily separated from noise, this is not the case after irradiation.
  If  signals from single Geiger discharges can be identified, the methods to characterise SiPMs are well established~\cite{Klanner:2018}.
  This however is not the case for a situation as shown for $\Phi = 10^{13}$~cm$^{-2}$.

  \begin{figure}[!ht]
   \centering
    \includegraphics[width=1.0\textwidth]{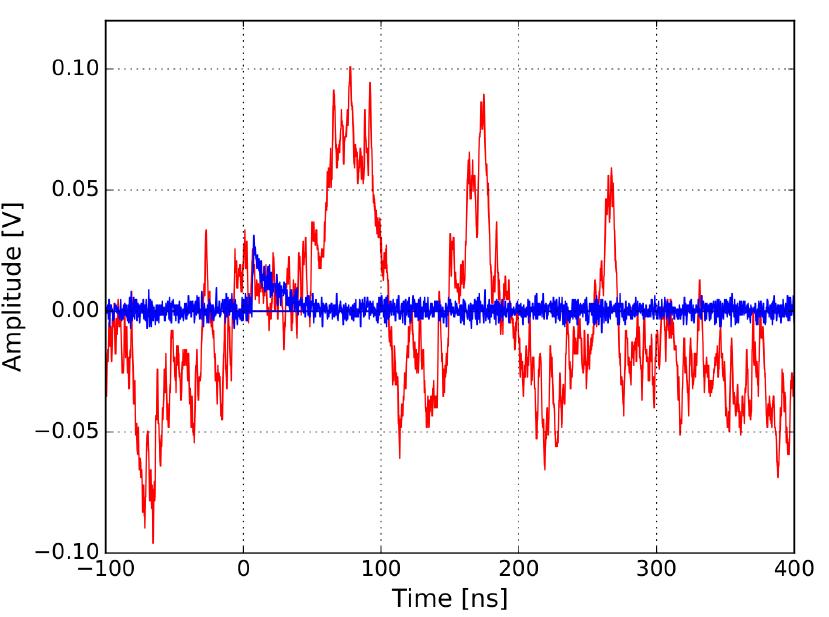}
    \caption{ SiPM current transients before (small signal -- blue trace) and after (erratic transient -- red trace) neutron irradiation to $10^{13}$~cm$^{-2}$.}
   \label{fig:Transient}
  \end{figure}
  \emph{Highly irradiated} is thus defined as the situation in which the signals from individual Geiger discharges cannot be separated due to a high $DCR$.
  Note that most methods proposed in this paper assume that after radiation damage the individual pixels of the SiPM show the same behaviour.
  In Ref.~\cite{Frach:2009, Engelmann:2018} it is shown that this is not the case before irradiation.

 \subsection{Doping density and electric field}
  \label{subsect:Doping}

  The question addressed is:
  Does the doping density, $N_d$, and the electric field change with fluence $\Phi$?
  To answer this question capacitance-voltage ($C-V$) measurements for reverse voltages between $V = 0.5 $ and 26~V have been taken at 10~kHz and the $AC$-voltage $V_{AC} = 0.5$~V.
  For the analysis the simplified model of an abrupt, one-dimensional $p^+ n$~junction is used:
  \begin{equation}\label{equ:Doping}
    x(V) = \frac{\varepsilon_{Si}~A} {C(V)} \hspace{3mm} \mathrm{and} \hspace{3mm} N_d(x) = \frac{2} {q_0~\varepsilon_{Si}~A^2} \cdot \frac{1} {\mathrm{d}(1/C^2)/\mathrm{d}V},
  \end{equation}
  where $x$ is the distance from the $p^+ n$~junction, $\varepsilon_{Si}$ the dielectric constant of silicon, $q_0$ the elementary charge and $A$ the total area of the SiPM.
  The validity of this simple method for this particular SiPM is demonstrated in Ref.~\cite{Chmill1:2017}, where it is shown that compatible $N_d(x)$ results are found for SiPMs with the same area $A = 1$~mm$^2$ and number of pixels between 1 and 4384.

  Up to $\Phi = 10^{13}$~cm$^{-2}$ no change in $N_d(x)$ is observed.
  At $\Phi = 5 \times 10^{14}$~cm$^{-2}$ a decrease by a few percent, as expected from donor removal, is found.
  It is concluded that up to  $\Phi=5 \times 10^{14}$~cm$^{-2} $ the change in doping density and thus of the electric field is a minor effect.

 \subsection{Electrical parameters}
  \label{subsect:ElParameters}

 The influence of radiation damage on the single-pixel capacitance, $C_{pix}$, the quenching resistor, $R_q$, and the capacitance parallel to the quenching resistor, $C_q$, has been investigated using admittance--frequency, $Y-f$ and for $R_q$ also forward current, $I-V_{forw}$, measurements.
 The $Y-f$~measurements were made at the reverse voltage $V = 26.5$~V just below $V_{bd}$ at $-30~^\circ $C, and at $+20~^\circ $C for 13 frequencies, $f$, between 100~Hz and 2~MHz.
 As discussed in Ref.~\cite{Klanner:2018}, from $Y(f)$ serial and parallel capacitances and resistances are obtained, which are then described by an electrical model of the SiPM.
 Each pixel is described by the capacitance of the diode, $C_{pix}$, in series with the parallel connection of $C_q$ and $R_q$.
 For the biasing lines an inductance is introduced, and for the dark current a parasitic resistance in parallel to the $N_{pix}$ pixels.
 The values of the model parameters are varied until the frequency dependence of the measured serial/parallel capacitances and resistances are reproduced.
 The model describes the data and the electrical parameters are well determined.
 For $C_q$ only an upper limit of $\approx 5$~fF is obtained.

   \begin{figure}[!ht]
   \centering
    \includegraphics[width=1.0\textwidth]{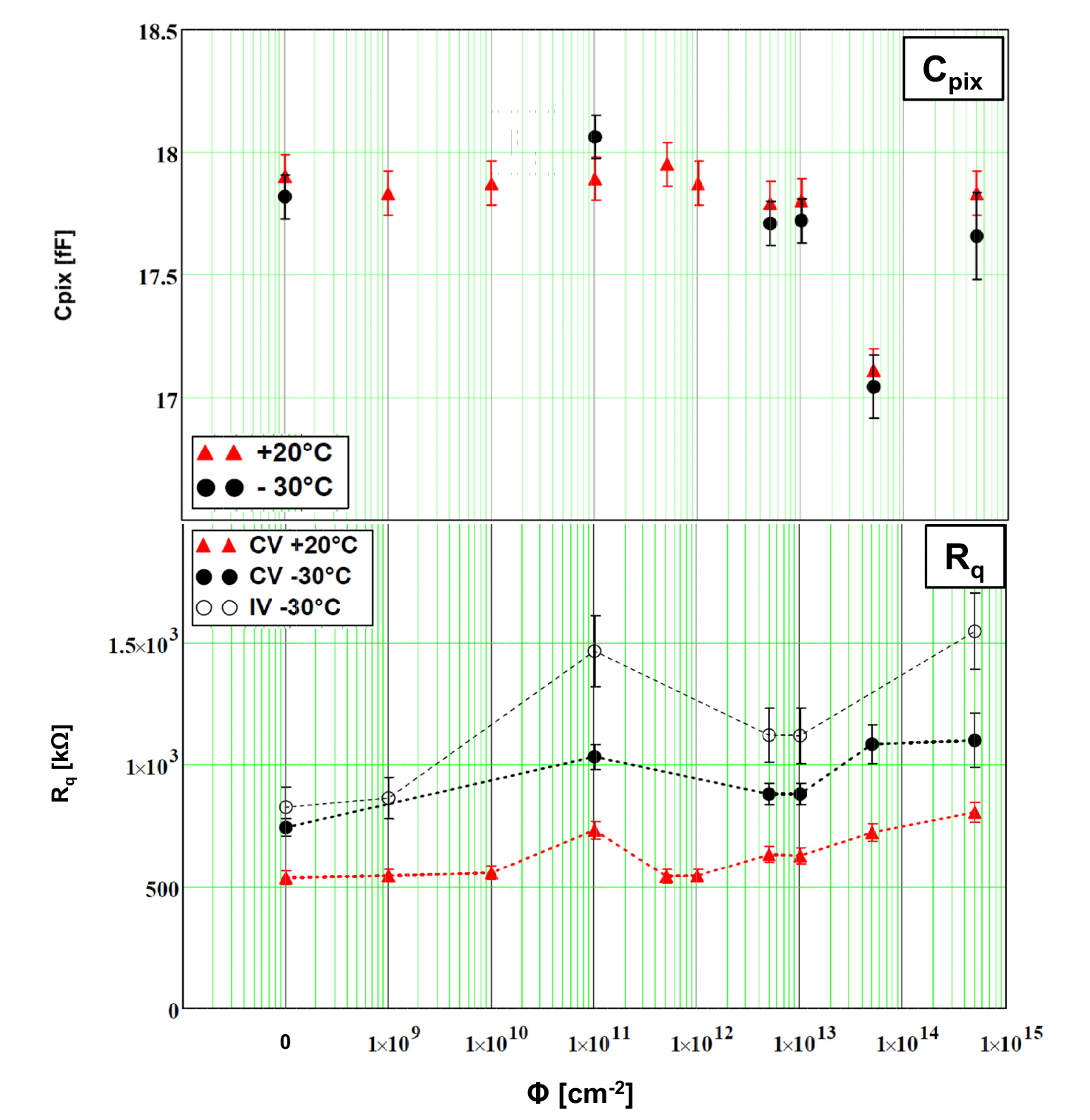}
    \caption{ Fluence dependence of $ C_{pix} $ from the $Y(f)$ (top), and of $R_q$ from the $Y(f)$ and $I(V_{forw})$ measurements (bottom).
    The full symbols are from the $Y(f)$ and the open symbols from the $I(V_{forw})$ data.}
   \label{fig:CpRq}
  \end{figure}
 Fig.~\ref{fig:CpRq}~top shows the results for $C_{pix}$.
 With the exception of the SiPM irradiated to $5 \times 10^{13}$~cm$^{-2}$, which shows some anomalies, $C_{pix} = 17.8 \pm 0.1$~fF independent of dose and temperature.
 As the SiPM gain $G \propto (C_{pix} + C_q)$, no change of $G$ with $\Phi $ is therefore expected from these results.
 Fig.~\ref{fig:CpRq}~bottom shows the results for $R_q$.
 The $Y-f$~results fluctuate at the $\pm 20$~\% level, which is ascribed to the technology of the poly-Si resistor.
 The comparison of the $- 30~^\circ $C and the $+ 20~^\circ $C results shows that $R_q$ decreases with temperature, which is a property of the poly-Si resistor.
 For $\Phi \gtrsim 10^{12}$~cm$^{-2}$ an increase of $R_q$ is observed.

 The open symbols represent the results of the $R_q$~determination using $I(V_{forw})$.
 It assumes that at sufficiently high $V_{forw}$~values the voltage drop over the diode hardly changes with $V_{forw}$, and d$I$/d$V_{forw} \approx N_{pix}/R_q$.
 However, in particular for the radiation-damaged SiPMs, straight-line fits to $I(V_{forw})$ for $V_{forw}$ between 1.2~V and 2.0~V show significant deviations from a linear behaviour.
 Thus the results from this method, which is the standard one, cannot be trusted and the $Y-f$~method is considered to be more reliable.

 \subsection{Breakdown voltage $V_{bd}$}
  \label{subsect:Vbd}

 The breakdown voltage, $V_{bd}$, is obtained from the minimum of the inverses logarithmic derivative
 $ILD = ( \mathrm{d} \ln (I) / \mathrm{d} V )^{-1}$ \cite{Klanner:2018}.
 Below as well as above $V_{bd}$ straight lines $ILD(V) = |V - V_{bd}|/n$ are observed, from which follows that $I(V) \propto |V - V_{bd}|^n $.
 The sign of $n$ is negative below and positive above $V_{bd}$.

   \begin{figure}[!ht]
   \centering
    \includegraphics[width=1.0\textwidth]{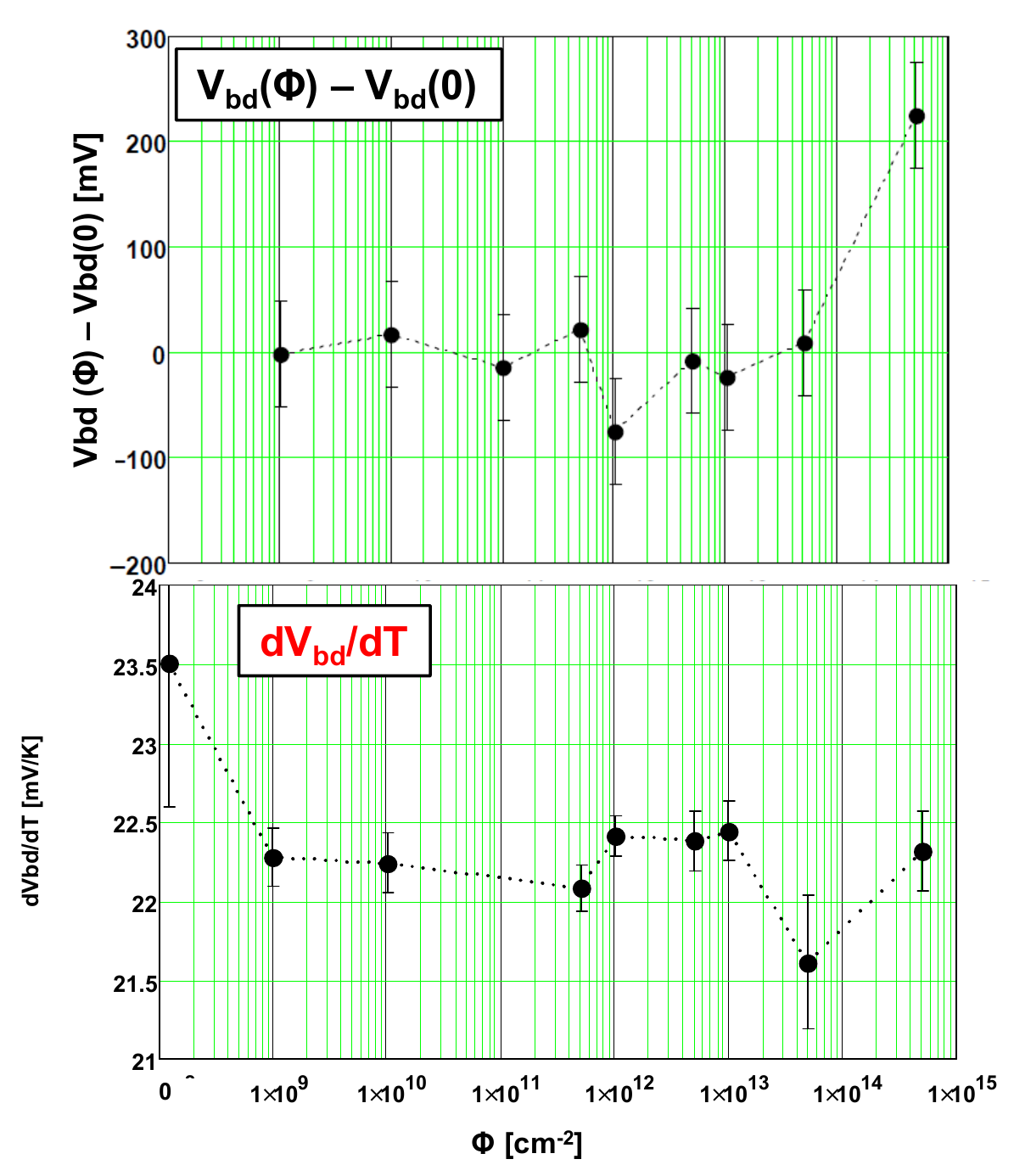}
    \caption{ Top: Difference $V_{bd}$ after and before irradiation measured at $20~^\circ $C.
    Bottom: Derivative d$V_{bd}$/d$T$ in the temperature range $-30~^\circ $C to $ + 30~^\circ $C as a function of neutron fluence.}
   \label{fig:Vbd}
  \end{figure}

 Fig.~\ref{fig:Vbd} top shows the difference of $V_{bd}$ after and before irradiation measured at $20~^\circ $C as a function of neutron fluence $\Phi$.
 Up to $\Phi = 5 \times 10^{13} $~cm$^{-2}$ $V_{bd}$ is independent of $\Phi$, whereas at $5 \times 10^{14} $~cm$^{-2}$ it  increases by $\approx 250$~mV.
 Note that the $\Phi $-dependence of $V_{bd}$ is expected to depend on the SiPM design.
 To investigate the $\Phi $-dependence of the change d$V_{bd}$/d$T$, $V_{bd}$ has been determined for the SiPMs after irradiation for 13 temperatures between $-30~^\circ $C to $30~^\circ $C.
 In all cases the temperature dependence of $V_{bd}$ can be described by straight lines.
 Fig.~\ref{fig:Vbd} bottom shows the results: d$V_{bd}$/d$T \approx 22.3 \pm 0.1 $~mV/$^\circ $C independent of $\Phi $.

 In Ref.~\cite{Chmill1:2017} it has been observed that $V_{off} $, the voltage at which the Geiger discharge turns off, differs from $V_{bd}$ for the SiPM investigated: $V_{bd} - V_{off} \approx 850$~mV.
 As the separation of individual Geiger discharges is required to determine $V_{off} $, we do not know how to determine it for highly-irradiated SiPMs.
 In the following analyses it is assumed that $V_{bd} - V_{off}$ does not change with $\Phi $.

 \subsection{Dose and temperature dependence of the dark current}
  \label{subsect:Idark}

\begin{figure}[!ht]
   \centering
    \includegraphics[width=1.0\textwidth]{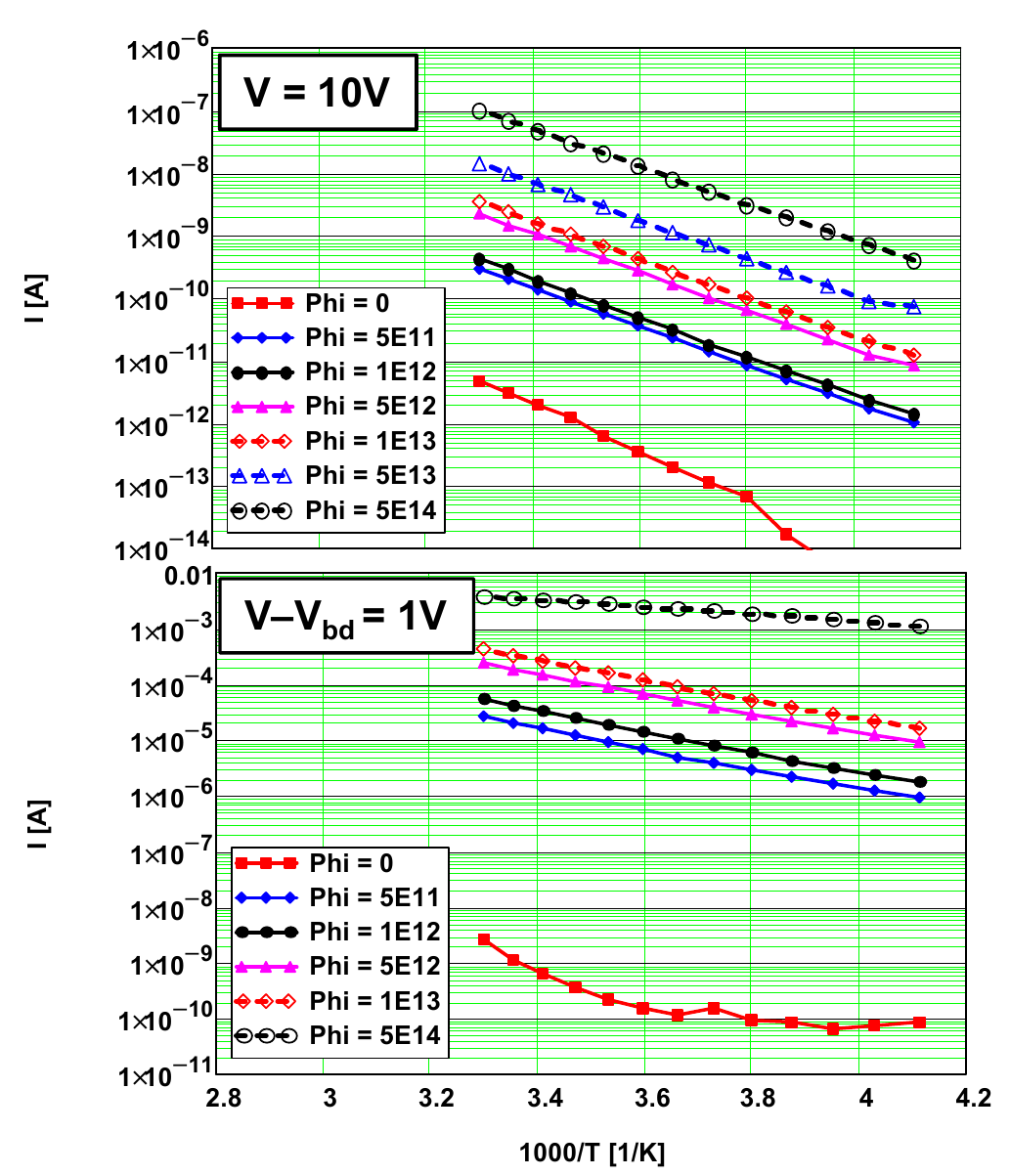}
    \caption{ Dark current as a function of $1000/T$ at $V = 10$~V (top), and at $V - V_{bd} = 1$~V (bottom).}
   \label{fig:IT}
  \end{figure}

\begin{figure}[!ht]
   \centering
    \includegraphics[width=1.0\textwidth]{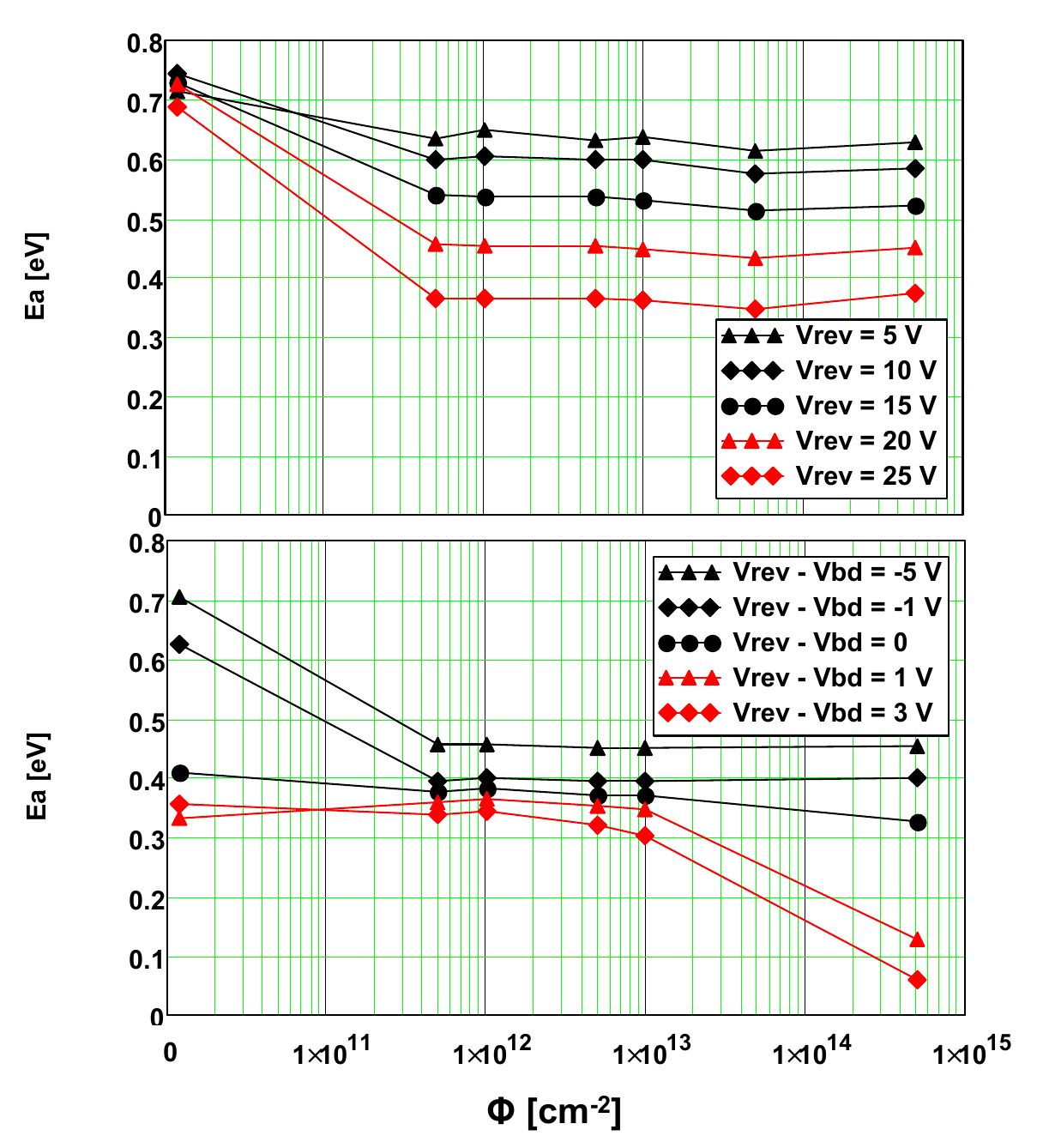}
    \caption{Arrhenius activation energies, $E_a$, as a function of $\Phi $ below $V_{bd}$ for different voltages (top), and for  different values of $V - V_{bd}$ (bottom). }
   \label{fig:Ea}
  \end{figure}

  From the temperature dependence of the dark current, $I_{dark} (T)$, an effective activation energy, $E_a$, can be determined  using the Arrhenius parametrisation $I_{dark} \propto e^{-E_a / (k_B \cdot T)}$, with $T$  the absolute temperature and $k_B$ the Boltzmann constant.
  Fig.~\ref{fig:IT} show Arrhenius plots for different $\Phi $~values at $V = 10$~V, which is below the amplification region, and for $V - V_{bd} = 1$~V.
  With the exception of the $\Phi = 0$ data for $V - V_{bd} = 1$~V, the Arrhenius parametrisation provides an adequate description of the data and $E_a$ can be determined.
  The results are shown in Fig.~\ref{fig:Ea}.

  For the non-irradiated SiPM for $V < V_{bd}$ a value $E_a \approx 0.7$~eV is found, which is compatible with the results from silicon sensors without amplification~\cite{Chilingarov:2013}.
  It should be noted that for the non-irradiated SiPM $I_{dark}$ is dominated by surface-generation.
  For $\Phi \geq 5 \times 10^{11}$~cm$^{-2}$, $E_a$ is independent of $\Phi $ with a value steadily decreasing from $\approx 0.62$~eV at 5~V to $\approx 0.35$~eV at 25~V, where charge amplification is important.
  A simple model using Shockley-Reed-Hall statistics for a single trap $t$, equal cross sections for electrons and holes, and taking into account the decrease of the band gap with $T$, gives for the energy difference of $t$ from the band center: $| E_t - E_i | = E_a - 0.65$~eV.
  Thus values $E_a \lesssim 0.65$~eV are nonphysical in this simple model.  
  For most of the data $E_a < 0.65$~eV is observed, and it is concluded that high-field effects are significant and the Arrhenius parametrisation describes the data, but does not determine the trap energies in the band gap.

  Close to $V_{bd}$, where charge multiplication is important, the temperature dependence of $V_{bd}(T)$ is taken into account by considering $I _{dark}(T)$ for fixed $V - V_{bd}$, as shown in Fig.~\ref{fig:IT} bottom for $I_{dark}(T)$ and in Fig.~\ref{fig:Ea} bottom for $E_a$.
  For $V - V_{bd} > 0$, where the SiPMs are used as photo-detectors, the decrease of $I_{datk} $ with $T$ at $\Phi = 5 \times 10^{14}$~cm$^{-2}$ is  slower than at lower $\Phi $~values:   $E_a = 0.5$\,eV corresponds to a decrease by a factor 2.2 for $\Delta T = 10$~K, whereas $E_a = 0.2 $\,eV to a factor 1.4.

 \subsection{Dark count rate}
  \label{subsect:DCR}

  Two methods are used to determine $DCR$: The dark-current and the variance method.
  For the relation between $I_{dark}$ and $DCR$ we assume:
  \begin{equation}\label{equ:IDCR}
    I_{dark} = DCR \cdot (C_{pix} + C_q) \cdot (V - V_{off}) \cdot ECF.
  \end{equation}
  Here $G = (C_{pix} + C_q) \cdot (V - V_{off})$ is the SiPM gain and $ECF$ the Excess Charge Factor, which describes the increase in signal due to cross-talk and after-pulsing.
  It has been obtained from charge measurements of the non-irradiated sensors~\cite{Chmill:2017, Vinogradov:2012}.

   \begin{figure}[!ht]
   \centering
    \includegraphics[width=1.0\textwidth]{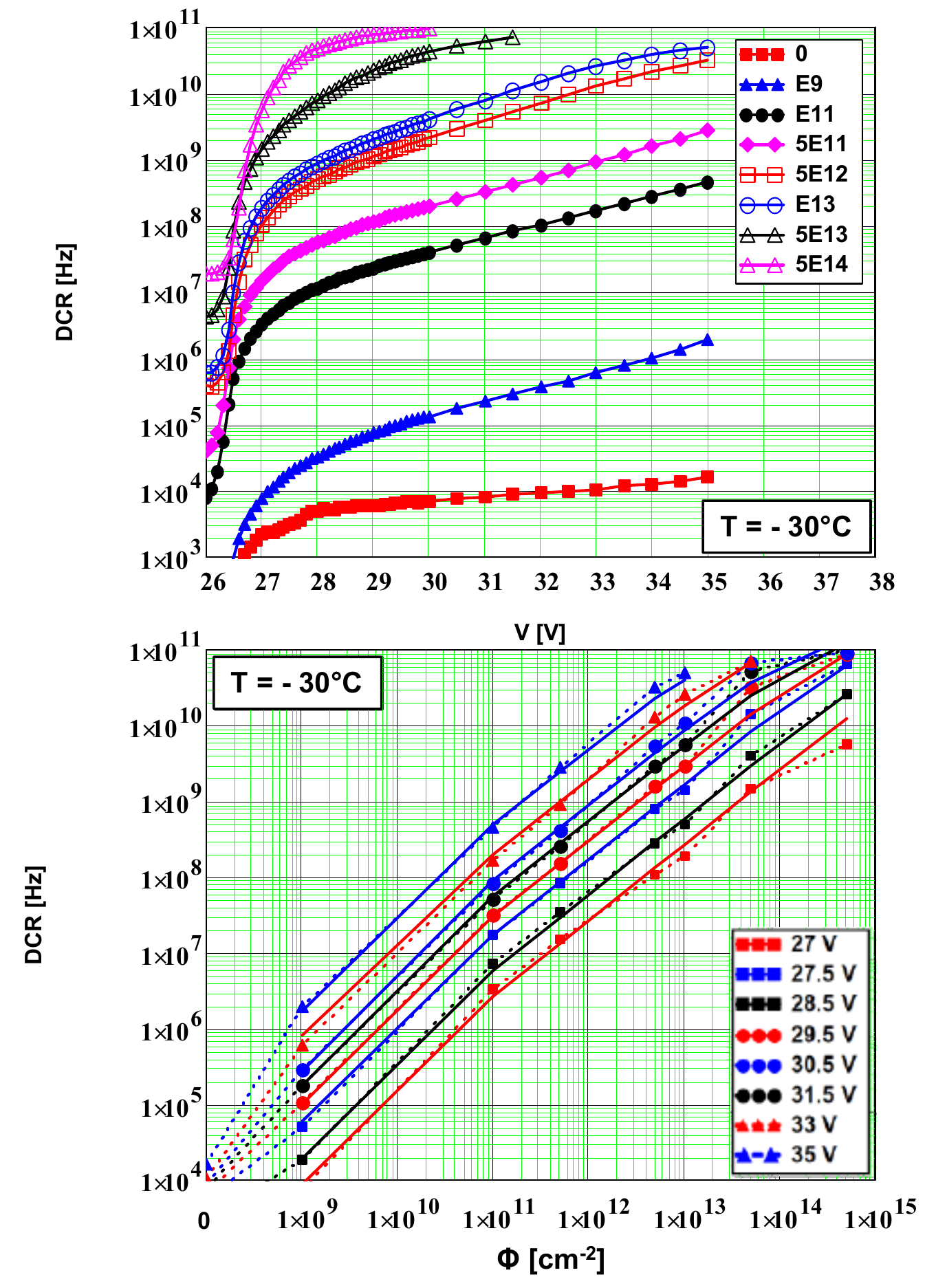}
    \caption{ Dark count rate, $DCR$, determined from the dark current at $- 30~^\circ$C,
    as a function of voltage for different $\Phi $~values (top), and
    as a function  of $\Phi $ for different voltages (bottom).
    The solid curves show the parametrisation discussed in the text. }
   \label{fig:DCR}
  \end{figure}

 Fig.~\ref{fig:DCR} shows the $DCR$ as a function of voltage and dose.
 An increase by 7 orders of magnitude is observed.
 A $DCR$~value of $10^{11} $~Hz corresponds to 100 dark counts per ns, and a mean pixel occupancy of 0.5 for the pixel recharging time $\tau \approx 20$~ns.
 Thus saturation effects, i.~e. dark counts, $DC$s, occurring when a pixel has not yet reached the bias voltage, are becoming significant.
 The relation between $DCR_0$, the $DCR$ without saturation, and $DCR_{sat}$ with saturation
 \begin{equation}\label{equ:DCRsat}
   \frac{DCR_{sat}} {DCR_0} = \frac{1} {1 + DCR_0 \cdot \tau / N_{pix} }
 \end{equation}
 can be derived for a SiPM transient with the shape $e^{-t / \tau}$.
 If we assume $DCR_0 (V,\Phi) = \alpha (V) \cdot \Phi + \beta (V)$, where $\beta (V)$ takes into account that $DCR (V)$ have different shapes for the non-irradiated and irradiated SiPM (see Fig.~\ref{fig:DCR} top), Eqs.~\ref{equ:IDCR} and \ref{equ:DCRsat} provide a parametrisation of $DCR (V, \Phi)$, which is shown as solid lines in Fig.~\ref{fig:DCR} bottom.
 A  model which takes into account the random arrival time of $DC$s, cross-talk and after-pulsing is under development.

 The second method uses the variance of the charge distribution, $Var_{dark}(t_{gate})$, obtained by integrating the current transients, as shown in Fig.~\ref{fig:Transient}, in different time intervals $t_{gate}$ before the signal from the sub-nanosecond laser.
 In Ref.~\cite{Klanner:2018} the following relation is derived:
 \begin{equation}\label{equ:VarDCE}
   DCR = \frac{Var_{dark}} {G^2 \cdot ENF \cdot ECF^2 \cdot \big(t_{gate} - \tau \cdot (1 - e^{-t_{gate}/ \tau } ) \big) },
 \end{equation}
 with the gain, $G $, and the excess noise factor, $ENF$, which has been determined with the non-irradiated SiPM~\cite{Chmill:2017, Vinogradov:2012}.
 For $10^9 \leq \Phi \leq 5 \times 10^{13} $~cm$^{-2}$ and $t_{gate}$ between 15 and 75~ns, the data are well described by Eq.~\ref{equ:VarDCE}, and the value of $\tau $ can be determined with an uncertainty of 1--2~ns, which is surprising for transients as shown in Fig.~\ref{fig:Transient}.
 For the non-irradiated SiPM, $Var_{dark}$ is dominated by electronics noise, and for $\Phi = 5 \times 10^{14}$~cm$^{-2}$ saturation effects become important, which are not accounted for in the model.

 For $10^{11} \leq \Phi \leq 10^{13}$~cm$^{-2}$ both methods agree within $\approx 20$~\%.
 We conclude that the dark-current method is simpler and also more reliable than the $Var_{dark}$~method.

 \subsection{Photon detection efficiency}
  \label{subsect:pde}

 In this section we address the questions:
 Up to which fluence and $DCR$ is the SiPM investigated still a useful photo-detector?
 Can we determine the loss in photon-detection efficiency, $PDE$, due to high pixel occupancies?
 For this study current transients, as shown in Fig.~\ref{fig:Transient}, with sub-nanosecond laser light resulting in $\approx 150$~Geiger discharges at $V - V_{bd} = 2$~V, were recorded, and integrated for a time $t_{gate}$ starting a few ns before the light signal.
 In the analysis the mean signal, $Mean_{light}$ and its variance $Var_{light}$ are used.
 Three methods are used to determine the number of photons initiating primary Geiger discharge, $\mu = N_\gamma \cdot PDE$, with $N_\gamma $ the number of photons hitting the SiPM:
 \begin{enumerate}
   \item $\mu = Mean _{light}/(G \cdot ECF)$,
   \item $\mu = Var _{light}/(G^2 \cdot ENF \cdot ECF^2$), and
   \item $\mu =Mean _{light} ^2 \cdot ENF / Var _{light}$.
 \end{enumerate}
 1. follows from the relation that the mean number of primary discharges is obtained from the ratio mean charge over gain with the correction $ECF$ for after-pulsing and cross-talk;
 2. can be derived in a similar way as Eq.~\ref{equ:VarDCE}; and
 3. uses the property of the Poisson distribution that its mean is equal to its variance.
 These relations can be rearranged to determine $G$.

 For the irradiated SiPMs up to fluences of $5 \times 10^{12}$~cm$^{-2}$ for the data at $- 30 ~^\circ$C, the three methods give consistent results.
 At higher $\Phi $-values differences, which increase with voltage, appear.
 It is expected that the differences are caused by saturation effects due to high $DCR$s, however an in-depth study has not  been made.
 Method 1. appears to be most straight-forward and reliable.
  \begin{figure}[!ht]
   \centering
    \includegraphics[width=1.0\textwidth]{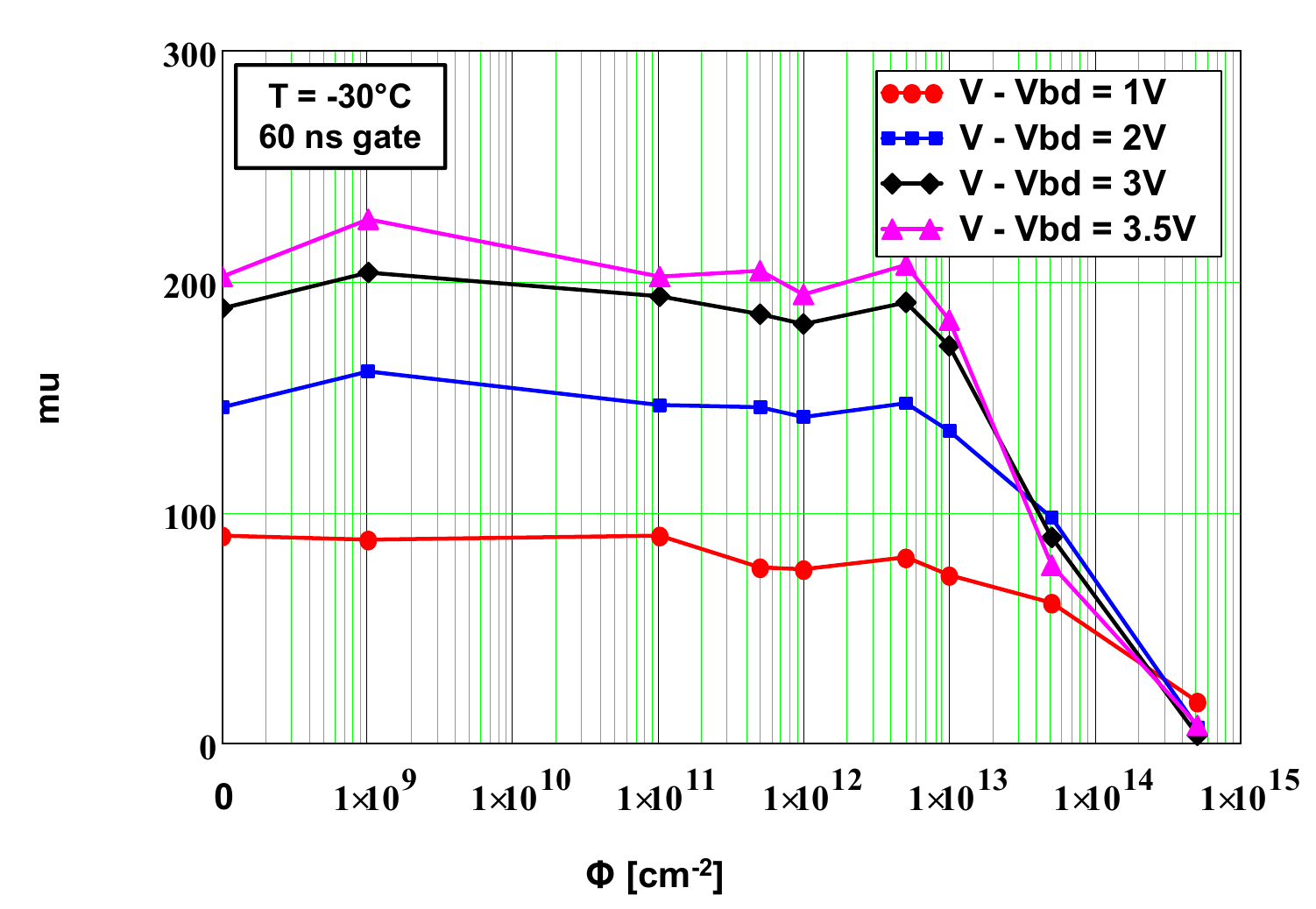}
    \caption{$\mu = N\gamma \cdot PDE(V,\Phi $) as a function of $\Phi $ for different values of $ V - V_{bd}$.}
   \label{fig:pde}
  \end{figure}

 In Fig.~\ref{fig:pde}, $\mu = N_\gamma \cdot PDE$ is shown as a function of $\Phi $ for $V - V_{bd}$  between 1~V and 3.5~V  at  $- 30~ ^\circ$C for $t_{gate} = 60$~ns.
 Up to $\Phi = 5 \times 10^{12}$~cm$^{-2}$, $\mu $ is independent of $\Phi $ within the uncertainties of $N_\gamma $.
 Above $\Phi = 5 \times 10^{12}$~cm$^{-2}$, $\mu $ decreases.
 As expected for saturation effects due to $DCR$, the relative decrease increases with bias voltage.
 From the figure it is concluded that, for the given measurement conditions, this particular SiPMs can be used as photo-detectors up to $\Phi \approx 5 \times 10^{13}$~cm$^{-2}$.
 The figure also shows that the operating range is larger at lower voltages.
 In addition, it can be shown that a shorter $t_{gate}$ or shorter pulse shaping allows to extend the maximum fluence up to which the SiPM can be operated.

  \section{Summary and conclusion}
  \label{sect:Coclusions}

 In this paper different methods to study highly radiation-damaged SiPMs are presented and used to characterise a prototype SiPM with 4384 pixels of $15 \times 15~\mu$m$^2$ area produced by KETEK.
 The SiPMs have been exposed to neutron fluences up to $\Phi = 5 \times 10^{14}$~cm$^{-2}$, and measurements were performed in the temperature range between $- 30~ ^\circ $C and $+30~^\circ $C.

 It is found that the effective doping density, $N_d$, and the pixel capacitance, $C_{pix}$, do not change with $\Phi $, whereas the poly-silicon quenching resistance, $R_q$, increases for $ \Phi \gtrsim 10^{13}$~cm$^{-2}$.
 The breakdown voltage, $V_{bd}$, remains constant up to $5 \times 10^{13}$~cm$^{-2}$, and then increases by $\approx 250$~mV, whereas d$V_{bd}$/d$T$ does not change with neutron irradiation.

 The dark current, $I_{dark}$, increases by 7 orders of magnitude with irradiation.
 From the temperature dependence of $I_{dark}$, the Arrhenius activation energies, $E_a$, are determined.
 It is found that for bias voltages above the breakdown voltages, $I_{dark}$ decreases by a factor of $\approx 2.2 $ for a temperature decrease by $10 ~ ^\circ$C, except  for $\Phi = 5 \times 10^{14}$~cm$^{-2}$, where this factor is $\approx 1.4$.
 Comparing the observed values of $E_a$ to the range allowed by Shockley-Reed-Hall statistics with additional simplifying assumptions, it is concluded that, due to high-field effects, $E_a$ can not be directly related to the energy of radiation-induced defects in the silicon band-gap.

 Two methods are used to determine the dark count rate, $DCR$, as a function of $\Phi $ and voltage.
 One uses $I_{dark}$, the other the variance of the charge distribution measured without illumination.
 Both methods give consistent results up to $DCR \approx 10^{10}$~Hz.
 For the highest fluences $DCR$ values of up to $ \approx 10^{11}$~Hz are observed, which corresponds to 100 dark counts per nanosecond.
 At such high $DCR$s the probability that a pixel is still recharging when the next Geiger discharge occurs, is significant, which results in saturation effects.
 A simple saturation model is proposed, which together with the observation that $DCR \propto \Phi$ before saturation, allows to parameterise in a simple way $DCR$ as a function of $\Phi $ and voltage.

 Finally, using charge spectra with the SiPM exposed to the light from a sub-nanosecond laser, the $\Phi $- and voltage-dependence of the photon detection efficiency, $PDE$, at $- 30~^\circ$C is estimated using three different methods.
 Up to $\Phi =  10^{13}$~cm$^{-2}$, $PDE$ is not affected by radiation damage within the measurement uncertainty of $\pm 10$~\%.
 For higher $\Phi $~values $PDE$ decreases rapidly and the SiPM is not anymore a useful photo-detector.
 At higher bias voltages the relative decrease in $PDE$ is larger and  sets in at lower $\Phi $~values.

 It should be noted that the methods proposed in this paper have been used for the analysis of a single SiPM type only, and thus may not be applicable to other types.
 The reader is strongly encouraged to use the different characterisation methods to find out to which extent they are useful and applicable for the SiPM investigated.

  \section*{Acknowledgements}

  We thank P.~Buhmann and M.~Matysek for keeping the equipment of the Hamburg Detector Laboratory in an excellent shape
  and E.~Fretwurst for enlightening discussions.
  Part of the measurements were performed by V.~Chmill, D.~Lomidze and M.~Nitschke.
  This project has received funding from the European Union's Horizon 2020 Research and Innovation programme under Grant Agreement no. 654168.

\end{document}